\documentclass[12pt,preprint]{emulateapj}

\newcommand{\etal}{{et~al.\null}}
\newcommand{\eg}{{e.g.,}}

\newcommand{\ie}{{i.e.,}}
\newcommand{\kms}{km~s$^{-1}$}

\newcommand{\simgt}{{\raise-.5ex\hbox{$\buildrel>\over\sim$}}}
\newcommand{\simlt}{{\raise-.5ex\hbox{$\buildrel<\over\sim$}}}
\newenvironment{packed_enum}{
\begin{enumerate}
  \setlength{\itemsep}{1pt}
  \setlength{\parskip}{0pt}
  \setlength{\parsep}{0pt}
}{\end{enumerate}}

\slugcomment{}

\shorttitle{Kinematic Evidence for Halo Substructure}
\shortauthors{Herrmann, Ciardullo, \& Sigurdsson}

\begin{document}

\title{Kinematic Evidence for Halo Substructure in Spiral Galaxies}

\author{Kimberly A. Herrmann\altaffilmark{1,2,3}, Robin Ciardullo\altaffilmark{1,2}, and Steinn Sigurdsson}
\affil{Department of Astronomy \& Astrophysics, The Pennsylvania State University \\ 525 Davey Lab, University Park, PA 16802}
\email{herrmann@lowell.edu, rbc@astro.psu.edu, steinn@astro.psu.edu}

\altaffiltext{1}{Visiting Astronomer, Cerro Tololo Inter-American Observatory. CTIO is operated by the Association of Universities for Research in Astronomy, Inc.\ (AURA) under contract to the National Science Foundation.}

\altaffiltext{2}{Visiting Astronomer, Kitt Peak National Observatory, National Optical Astronomy Observatories, which is operated by AURA under cooperative agreement with the National Science Foundation.  The WIYN Observatory is a joint facility of the University of Wisconsin-Madison, Indiana University, Yale University, and the National Optical Astronomy Observatories.}

\altaffiltext{3}{Current address: Lowell Observatory, 1400 West Mars Hill Road, Flagstaff, AZ, 86001}

\begin{abstract}
We present the results of a kinematic study of planetary nebulae in the extreme outskirts of two spiral galaxies, M83 (NGC~5236) and M94 (NGC~4736).  We find that in the inner regions of the galaxies, the vertical velocity dispersion ($\sigma_z$) falls off exponentially with the light, as expected for a constant mass-to-light ratio, constant thickness disk.  However, starting at four optical scale lengths, $\sigma_z$ asymptotes out at roughly 20~\kms.  Our analysis finds evidence for significant flaring in the outer regions as well, especially in M94.  These observations are in excellent agreement with predictions derived from models of disk heating by halo substructure, and demonstrate how kinematic surveys in the outer disks of spirals can be used to test hierarchical models of galaxy formation.
\end{abstract}


\section{INTRODUCTION}
The cold dark matter (CDM) paradigm has proven to be very successful in explaining the large-scale structure of the universe and galactic clusters \citep{t+04}, yet problems still exist at the galactic level.  One such issue is the ``missing satellite'' problem:  according to numerical simulations, the structure of a galactic halo should look like a small version of a galaxy cluster and contain many small subhalos \citep[\eg][]{m+99,k+99}.  The recent discoveries of satellite galaxies tidally stripped by the Milky Way \citep[][and references therein]{i+94,n+02,j+08} and Andromeda \citep{f+05,i+07} help to solve this problem.   These small companions have a profound effect on the morphology and kinematics of thin galactic disks.  Specifically, numerical models of satellite halo bombardments predict

\begin{packed_enum}
\item the formation of faint stellar streams above the disk plane (which would be very difficult to observe in any but the nearest galaxies),
\item the generation of long-lived, low-surface brightness, dynamically cold, ring-like features in outer disks,
\item the growth of a strong bar,
\item the production of a pronounced flare, and
\item the development of a thick disk
\end{packed_enum}
\citep[][and references therein]{kb+08}.  Though many simulations \citep[][and references therein]{qhl93,f+01,hc06,kb+08} have been performed to explore disk heating by halo substructure, observationally the results from this process have proven to be rather elusive.  Only recently have teams begun to detect signs of this process in the Andromeda Galaxy \citep{r+08} and other nearby galaxies \citep{dJ+08}.

Here we describe kinematic evidence for halo substructure derived from the velocity dispersion of planetary nebulae (PNe) in the extreme outer disks of two nearby, face-on spirals.   We use stability arguments to show that the disks of M83 (NGC~5236) and M94 (NGC~4736) must flare at radii greater than $\sim$4~scale lengths and demonstrate that at these large radii, the $z$ velocity dispersions of the old disk stars agree with the results of numerical simulations of a disk heated by halo substructure \citep{hc06}.

\section{THE SURVEY}
Flat rotation curves indicate the presence of dark matter in the outer regions of spiral galaxies and allow us to determine total galactic mass \citep[\eg][]{sr01}.  However, rotation curves alone cannot decouple the mass contribution of the disk from that of the dark halo \citep{bsk04}.  To break this degeneracy, we have been using the motions of planetary nebulae in low-inclination spirals to measure disk mass directly via the $z$~motions of stars.  PNe are ideal particles for this purpose: they are bright, abundant to $>$5 scale lengths, representative of the old disk, relatively easy to distinguish from \ion{H}{2} regions (via their distinctive [\ion{O}{3}]-H$\alpha$ ratio; Ciardullo \etal\ 2002), and amenable to precise ($\sim$2 \kms) radial velocity measurements with fiber-fed spectrographs.

In \citet{thesis1} (Paper~I), we presented the results of a narrow-band imaging survey of six low-inclination nearby spirals (IC~342, M74 (NGC~628), M83 (NGC~5236), M94 (NGC~4736), NGC~5068, and NGC~6946) in which we identified 165, 153, 241, 150, 19, and 71 PNe, respectively.  In two upcoming papers (K.A.\ Herrmann \& R.\ Ciardullo 2009a, 2009b, in preparation, Papers II and III), we will detail our spectroscopic follow-up observations, present high-precision radial velocities for 550 of our PN candidates, and use the data to estimate dynamical disk masses.

Here, we focus on the data set for the two galaxies for which we have the largest radial coverage, the SBc spiral M83, which is at a distance of 4.8~Mpc, and the earlier Sab system M94 at $D = 4.4$~Mpc (Paper~I).  Our PN velocity sample for the former galaxy consists of 162 objects at radii between 2.5 and 24.7~kpc, \ie\ between $\sim$1 and 10 optical disk scale lengths.  These data, which were taken with the Hydra bench spectrograph on the Cerro Tololo Inter-American Observatory (CTIO) 4m telescope, have typical velocity uncertainties of $\sigma_v \sim 6.5$~\kms, and in all cases, $\sigma_v < 15$~\kms.  For M94, our data set contains 127 planetaries observed with the Hydra spectrograph on the WIYN (Wisconsin, Indiana, Yale \& NOAO) telescope.  These PNe have galactocentric radii between 0.5 and 8.8~kpc ($0.4 < R < 7$~disk scale lengths), and typical velocity uncertainties of $\sigma_v \sim 4$~\kms, again with $\sigma_v < 15$~\kms\ for all objects (see Paper~II).

Since neither galaxy is exactly face-on, we began our analysis by removing the effects of galactic rotation from the PN sample.  This was done using velocity maps from The \ion{H}{1} Nearby Galaxy Survey \citep[THINGS;][]{THINGS} with a correction for asymmetric drift.   We then binned the planetaries by radius (with $15-16$~PNe per bin in M83, and $17-18$~PNe per bin in M94), and identified those objects more than $\sim$2.5~$\sigma$ away from their bin mean as possible halo contaminants.  (In practice, this made very little difference to the analysis, since the procedure eliminated only six objects in M83 and three in M94.)   Finally, to extract the component of the velocity dispersion perpendicular to the galactic disk, $\sigma_z$, from the other two constituents of the velocity ellipsoid, we began by using the epicyclic approximation for near-circular orbits to couple $\sigma_{\phi}$ to $\sigma_R$.   We then constrained the shape of the disk velocity ellipsoid using the limits imposed by the physics of disk scattering \citep[$\sigma_z < \sigma_R$;][]{v85, jb90} and bending instabilities \citep[$\sigma_z > 0.25 \sigma_R$;][]{t66, a85, ms94}, and computed the likelihood that each combination of $\sigma_z$ and $\sigma_R$ could produce the line-of-sight dispersions observed within our bins (see Paper~III).

\begin{figure}
\epsscale{1.0}
\plotone{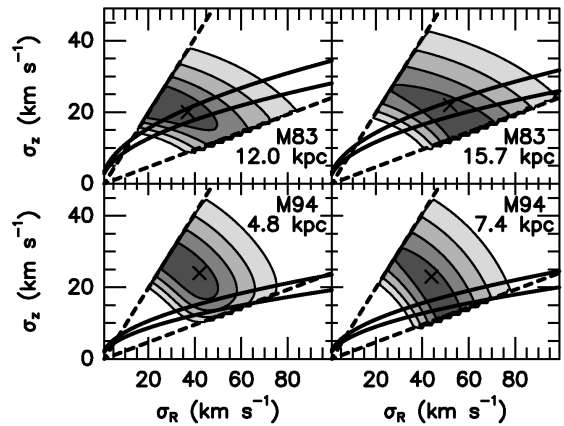}
\caption{Maximum likelihood probability contours for the outer disks of M83 and M94.  From dark to light, the contour regions enclose 38\% (0.5$\sigma$), 68\% (1$\sigma$), 86\% (1.5$\sigma$), and 95\% (2$\sigma$) of the probability.  The dashed lines show the limits of our analysis; the black crosses indicate the most likely solutions.  The solid curves display the upper limits on $\sigma_z$ derived from the additional constraint that the disk be stable against axisymmetric perturbations.  The multiple curves represent intermediate (sech ($z$)) disks with scale heights of 200 and 300~pc.  Note that for stability, $h_z$ must be larger than these nominal values, especially in M94.
\smallskip
 \label{contours} }
\end{figure}

Figure~\ref{contours} shows the results of this analysis, displaying contours which enclose 38\% ($0.5 \sigma$), 68\% ($1 \sigma$), 86\% ($1.5 \sigma$), and 95\% ($2 \sigma$) of the probability.  Note that these contours have a slight tilt to them.  This is a consequence of the galactic inclination:  neither galaxy is precisely face-on ($i \sim 24^\circ$ for M83, $i \sim 35^\circ$ for M94), and the larger the inclination, the more difficult it is for our maximum likelihood procedure to extract $\sigma_z$ from the line-of-sight velocity dispersion.  Nevertheless, the figure does demonstrate that our measurements of $\sigma_z$ are reasonably well defined, especially for M83.

\section{RESULTS}
\begin{figure}
\epsscale{1.0}
\plotone{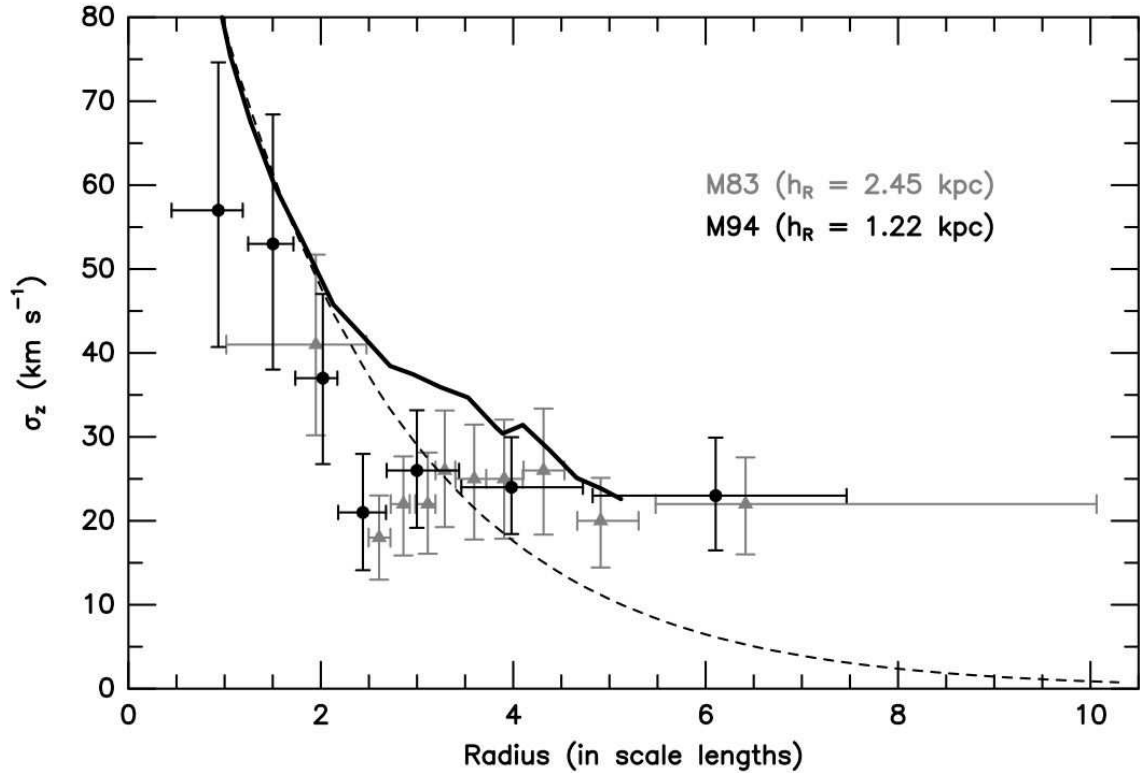}
\caption{Comparison of our M83 and M94 results to the $\sigma_z$ values (solid black curve) determined from the Model~F ($h_R$ = 3.5 kpc) numerical simulation of \citet{hc06}.  The data have not been rescaled in any way.  The agreement between the data and model is much better than with the constant mass-to-light ratio, constant scale-height exponential disk (dashed curve).
\label{subhalo} }
\end{figure}

For spiral galaxies, the velocity dispersion perpendicular to the disk is related to disk surface mass, $\Sigma(R)$, and the disk scale height, $h_z$, by
\begin{equation}
\sigma_z^2(R) = K G \Sigma(R) h_z,
\label{isothermal}
\end{equation}
where $K = 2 \pi$ for the isothermal case, $K = \pi$ for disks whose vertical mass density drops exponentially, and $K = \pi^2/2$ for the intermediate sech($z$) case \citep{vdK88}.  If (1) the disk mass-to-light ratio is constant, (2) the disk light of the galaxy decays exponentially with a single scale length ($h_R$), and (3) the disk scale height is also constant, then $\sigma_z$ should decrease exponentially, with a scale length twice that of the light.  As the marginalized probabilities of Figure~\ref{subhalo} illustrate, this is not the case in either galaxy.  While $\sigma_z$ does track the light in the galaxies' inner regions, the curve flattens out at distances more than four disk scale lengths from the nucleus.  The minimum dispersion of $\sigma_z \sim 20$~\kms\ is much greater than the typical measurement error of $\lesssim$5~\kms, and greater than the $\lesssim$15~\kms\ values expected from the velocity dispersion of ionized gas and \ion{H}{2} regions \citep[\ie][]{zhe90, f+07}.  Moreover, it is difficult to conceive of any way that internal extinction could create such an effect.  While it is true that nonisothermal disks may have velocity dispersions that increase with height above the plane \citep[see][]{vdK88}, any layer of dust that is thick enough to effect our derived values of $\sigma_z$  would also extinct the bulk of the PN population below the detection threshold and alter the distribution of PN [\ion{O}{3}]/H$\alpha$ line ratios (see Paper~II).  Finally, large-scale warping of the disks is not a solution: even with modeling isophotal twists by adjusting the disk position angle and inclination 	by $\sim$70$^\circ$ and $\sim$15$^\circ$, respectively, we cannot produce the dispersion curve seen in  Figure~\ref{subhalo} (see Paper~III).

The flat velocity dispersion profile of M94 is partially explained by its unique 3.6~$\mu$m-band radial profile, which has one scale length ($h_R = 1.22$~kpc) in its inner regions, then breaks to follow a shallower profile ($h_R = 7.16$~kpc) at radii greater than 5~kpc \citep{dB+08}.  (A recent deep $R$-band profile shows the same broken shape (Erwin 2009, private communication).)  But M83's disk has no such break, as its simple exponential profile ($h_R = 2.45$~kpc) extends over $\sim$20~kpc to the edge of our survey area \citep{dJ+08}.  Moreover, even with its broken disk profile, M94's $\sigma_z$ values are much higher than possible for a constant $M/L$, constant $h_z$ disk.  

These high $\sigma_z$ values also bring up another issue.  To be stable against axisymmetric perturbations, a thin stellar disk must obey the \citet{t64} criterion
\begin{equation}
\sigma_R > {3.36 G \, \Sigma \over \kappa},
\label{toomre}
\end{equation}
where $\kappa$ is the epicyclic frequency of the orbits.  Combining this criterion with Equation~(\ref{isothermal}) yields a constraint on the allowable values of $\sigma_z$:
\begin{equation}
\sigma_z < \left( {K h_z \kappa \sigma_R \over 3.36} \right)^{1/2}.
\end{equation}
Because our galaxies are low-inclination systems, $h_z$ is not directly obtainable from our observations.  However, imaging surveys of large, edge-on, noninteracting spirals demonstrate that the scale length to scale height ratio of galactic disks ranges from $\sim$10 in late-type systems to $\sim$5 in earlier-type objects \citep[\eg][]{dG98, kvd02}.  When applied to the two galaxies studied here, this relation implies that $h_z$ should be between 200 and 300~pc for both the Sab spiral M94 \citep[$h_R = 1.22$~kpc;][]{dB+08} and the SBc system M83 \citep[$h_R = 2.45$~kpc;][]{dJ+08}.   Yet, as the curves of Figure~\ref{contours} indicate, outer disks as thin as this are unlikely, as our solutions for $\sigma_z$ in M83 and especially M94 lie predominantly above the stability limit.  This suggests that the stellar disks of these systems flare dramatically in their outer regions.

Could this flaring and higher than expected values of $\sigma_z$ be due to heating of the disk by halo substructure?  We can address this question by comparing our results to $N$-body simulations of dark subhalo interactions with large galaxies.  In particular, \citet{hc06} have modeled the heating of a Milky Way-like galaxy by a population of subhalos, distributed according to a \citet{h90} law.  Figure~\ref{subhalo} compares our $\sigma_z$ velocity dispersion profile to their Model~F, which reproduces interactions between an initially thin, stable, constant mass-to-light ratio disk, and a few ($\sim$300) massive ($\lesssim$10$^9 M_{\odot}$) subhalos, totaling $M_{sh} \lesssim 0.15 M_{disk}$ and distributed with a half-mass radius of $\sim$210~kpc.  Although their model only probes out to $\sim$5~disk scale lengths, the agreement between observations and theory is excellent.  Their model not only reproduces the run of $\sigma_z$ versus radius seen in our galaxies, but also predicts an amount of disk flaring that is consistent with our analysis.  Moreover, \citet{s+08} have recently reported that, of the 23 known Milky Way satellite galaxies, the 18 with dynamical mass measurements all have masses of $M \sim 10^7 M_{\odot}$ within a radius of 0.3~kpc.  This suggests that these satellites once had total masses of $\sim$10$^9 M_{\odot}$ before encountering the Milky Way's potential \citep{s+08}.  This value is in excellent agreement with the subhalo masses used in \citet{hc06}, and supports the idea that interactions with subhalos could be responsible for the flat dispersion profiles seen in the outskirts of galaxies.

\section{CONCLUSIONS} 
The existence of persistent very thin disks in spiral galaxies has long been recognized to be a significant constraint for models of structure growth \citep[cf.\null][]{m+99}.  Major mergers destroy thin disks, while large asymmetries or substructure in extended dark matter halos will warp or heat these structures \citep{qhl93, f+01}.  Conversely, evidence for extended thick disks (or thin young disks embedded within older, thicker disks) can preserve direct evidence of past mergers and/or episodes of externally triggered star formation.

The two galaxies considered here have very different radial profiles: while the slope of M94's ``antitruncated'' stellar disk changes dramatically at a radius of $\sim$5~kpc \citep{dB+08}, M83's exponential profile appears undisturbed out to $\sim$10~disk scale lengths \citep{dJ+08}.  Thus, while M94 may have been involved in a minor merger some years ago \citep{y+07}, the only evidence for interaction in M83 is the warping of its extreme outer \ion{H}{1} disk, which begins $\sim$8~scale lengths from the nucleus \citep{M83HI}.  Yet in both these systems, the kinematic evidence suggests that the stellar disk flares dramatically past $\sim$4~disk scale lengths, and that at these large distances, the $z$~velocity dispersion is independent of radius.  This result is consistent with cosmological models of hierarchical structure formation, where a thin disk is heated by a relatively small number of massive subhalos, which are either embedded within the parent dark halo with a relatively small initial spatial spread \citep{hc06}, or initially on radial orbits \citep{h+08}.  Since simulations indicate the disk heating does not scale linearly with the mass and number of subhalos, this is potentially a strong constraint on substructure.  While there are several different processes that may combine to heat a disk, cooling such a disk back to a low velocity dispersion state is difficult.   (The slow accretion of extremely large amounts of cold gas could, in theory, lead to adiabatic compression of the disk.  However, in such a process, the gas mass would need to be many times that of the stellar disk, and even then, there would be stability issues.)  Our current measurements of $\sigma_z$ at large radii already rule out some models for halo substructure, and are directly consistent with models where there are fewer, more massive subhalos on orbits with relatively low angular momentum.

More data are required to explore this constraint further.  In particular, measurements over a wider range of galaxy masses, radii, and Hubble types are needed.  Fortunately, such data are relatively straightforward to obtain with moderate-sized telescopes.  Planetary nebula surveys in the extreme outer disks of the low-inclination, late-type spirals M74, M101, and IC~342 are already under way, and observations in earlier systems, such as M95 and NGC~1291 are possible.  Thus, a robust comparison to hierarchical models of structure formation should be possible in just a few years.

\acknowledgments
We thank Fabian Walter and Erwin de Blok for giving us access to the THINGS data prior to publication.  We also thank our anonymous referee for useful comments.  This research has made use of NASA's Astrophysics Data System and the NASA/IPAC Extragalactic Database (NED) which is operated by the Jet Propulsion Laboratory, California Institute of Technology, under contract with the National Aeronautics and Space Administration.  This work was supported by NSF grant AST 06-07416 and a Pennsylvania Space Grant Fellowship.

\end{document}